\documentclass[reprint,
amsmath,amssymb,
aps,
]{revtex4-2}

\usepackage{graphicx}   
\usepackage{dcolumn}    
\usepackage{bm}         
\usepackage{siunitx}    
\usepackage{xcolor}    

\newcommand{\BXf}{B_{\mbox{\scriptsize Xf}}}
\newcommand{\muPol}{\mu_{\mbox{\scriptsize Pol}}}

\newcommand{\JXf}{J_{\mbox{\scriptsize Xf}}\,}
\newcommand{\JXfS}{J_{\mbox{\scriptsize Xf}}S\,}
\newcommand{\RPol}{R_{\mbox{\scriptsize Pol}}}
\newcommand{\aB}{a_{\mbox{\tiny B}}}
\newcommand{\BSAT}{B_{\mbox{\tiny SAT}}}
\newcommand{\muB}{\mu_{\mbox{\tiny B}}}
\newcommand{\mueff}{\mu_{\mbox{\tiny eff}}}
\newcommand{\ocite}[1]{\onlinecite{#1}}
\newcommand{\kB}{k_{\mbox{\tiny B}}}
\newcommand{\TC}{T_{\mbox{\tiny C}}}
\newcommand{\TN}{T_{\mbox{\tiny N}}}
\newcommand{\gS}{g_{\mbox{\tiny S}}}
\newcommand{\MSAT}{M_{\mbox{\tiny SAT}}}
\newcommand{\Brillouin}{B_{\mbox{\scriptsize S}}}

\begin{document}

\preprint{APS/123-QED}

\title{Modeling huge photoinduced spin polarons in intrinsic magnetic semiconductors}

\author{S. C. P. van Kooten}
\affiliation{Instituto de Fisica, Universidade de Sao Paulo, 05315-970 Sao Paulo, Brazil}

\author{X. Gratens}%
\affiliation{Instituto de Fisica, Universidade de Sao Paulo, 05315-970 Sao Paulo, Brazil}

\author{A. B. Henriques}%
\email{andreh@if.usp.br}
\affiliation{Instituto de Fisica, Universidade de Sao Paulo, 05315-970 Sao Paulo, Brazil}

\date{\today}

\begin{abstract}
In intrinsic magnetic semiconductors, the absorption of a single photon can generate a spin polaron, whose magnetic moment reaches many thousands of Bohr magnetons  \cite{prl2018,apl2020}. Here we investigate photoinduced spin polarons, using Monte Carlo simulations. In antiferromagnetic semiconductors, photoinduced spin polarons are most efficiently generated in the whole temperature interval up to the phase transition, whereas in ferromagnetic semiconductors much larger spin polarons can be photoinduced, but only around the phase transition temperature. Because Monte Carlo simulations are computationally expensive, we developed an analytical model, based on the Weiss field theory. Although the Weiss model does not provide as much information as a Monte Carlo simulation, such as spin texture and fluctuations, it yields formulas that can be used to estimate instantly the expected photoinduced spin polaron size in many intrinsic magnetic semiconductors.
\end{abstract}


\maketitle


\section{Introduction}
\label{sec:intro}


Recently, it was experimentally  demonstrated that in the antiferromagnetic members of the europium chalcogenide series EuX (X=O, S, Se, Te), a single photon can generate a very large spin polaron (SPs), reaching hundreds, or even thousands of Bohr magnetons \cite{prb08,apl2011,prb2014,prb16Rapid,prb17EuTe,prl2018}.  In the ferromagnetic compound EuS, the size of the SPs surpasses 10,000 Bohr magnetons \cite{apl2020}. These extraordinarily large photogenerated SPs can be used to manipulate the magnetization on the picosecond time scale \cite{prl2018}, which is a topic of vast current interest \cite{FastestMagRecnature2004,Hellman_RevModPhys2017,kirilyukRMP}. SPs are formed when an incident photon generates an electron-hole pair, and in a few tens of picoseconds the lattice surrounding the excited electron becomes magnetized, due to the activation of the exchange interaction between the photoexcited electron and the lattice spins \cite{deGennes}.

This paper presents the first theory ever reported on the temperature dependence of size of very large photoinduced SPs in magnetic semiconductors. The work developed here is based on an effective mass self-consistent Hamiltonian describing the photoinduced SP, which includes the kinetic energy of the photogenerated band electron, its Coulomb interaction with  the hole, its exchange interaction with lattice spins, the exchange interaction between lattice spins, and an applied magnetic field \cite{prb2014}. The spin polarons induced by charge carriers can be described more fully by a Kondo Hamiltonian \cite{Horsch,Varma}, but this more complex approach is beyond the scope of this paper, given that the simple effective mass Hamiltonian \cite{prb2014} describes sufficiently well the temperature dependence of the SP size analyzed in this work.
We present results of Monte Carlo (MC) simulations, which reproduce very well available experimental data for all europium chalcogenides. However, the Monte Carlo simulations are computationally intensive, so we developed a simpler model, based on the Weiss field (WF) theory. The WF model, whose simplicity makes it accessible to a wide audience, produces analytical formulas that give the SP magnetic moment versus temperature in fair agreement with the MC simulations. These formulas can be extended to many other magnetic semiconductors.

The analytical formulas we obtained for the temperature dependence of the magnetic moment of the SP, dependent on few basic material parameters, are of crucial and practical value for researchers in this field, both in fundamental as well as applied physics.


\section{Spin Polaron modeling}
SPs are formed due to the exchange interaction between between an electron in the photoexcited state state and the lattice spins. This interaction is described by an effective magnetic field, $\BXf$, acting on the lattice spins, which at a distance $r$ from the electron is given by \cite{prb2014}
\begin{equation}
\BXf(r)=\frac{\JXfS}{N \mu^\ast}\Psi^2(r),
\label{eq:bxf}
\end{equation}
where X and f represent the excited electron and lattice spin, respectively, $\JXf$ is their exchange interaction,
$N$ and $S$ is the volume density and magnetic quantum number of lattice spins, respectively, $\mu^\ast=\gS\muB S$, $\gS$ is the Land\'e factor, $\muB$ is the Bohr magneton,  $\Psi(r)=\frac{e^{-r/\aB}}{\sqrt{\pi \aB^3}}$ is the Bohr wavefunction of the photoexcited electron, and $\aB$ is the effective Bohr radius. For EuX, $\gS=2$, $S=7/2$, $N=4/a^3$, $a$ is the face centered cubic lattice parameter. The photoexcited hole is strongly localized and immobile \cite{cho,prb09}, which renders negligible its exchange interaction with surrounding lattice spins.

Since the effective exchange field, given by eq.~\eqref{eq:bxf}, is non-zero for any $r$, one might naively assume that the photoexcited electron always causes some spin polarization of any lattice spin, no matter how distant the lattice spin is from the SP center. In reality, the photoexcited electron only polarizes lattice spins within a sphere of radius $\RPol$, where its exchange field wins the competition over other interactions affecting spin orientation. Magnetic anisotropy, lattice spin fluctuations, or an applied magnetic field are examples of competing interactions that will limit the radius of the SP. If we characterize the competing interactions by an effective magnetic field $B_0$, then $\BXf(r) \ge B_0$ delimits the SP sphere, of radius $\RPol$, which from equation  \eqref{eq:bxf} is equal to
\begin{equation}
\RPol=\frac{\aB}{2}\ln\left(\frac{\JXfS}{\pi\aB^3 N\mu^\ast B_0}\right).
\label{eq:RPol}
\end{equation}
This equation shows that the greater the competing field $B_0$, the smaller the radius of the SP.

In EuTe, the magnetic anisotropy is described by a spin-flop field of magnitude about $0.1$~T \cite{battleseverett,prb2014}, hence it is reasonable to assume $B_0=0.1$~T for EuTe.
Using other EuTe parameters, given in Table~\ref{tab:pars},
equation \eqref{eq:RPol} gives $\frac{\RPol}{a}=4.2$, in agreement with experimental measurements \cite{apl2011}. As shown in Ref. [\onlinecite{prb2014}], $\RPol$ is nearly independent of the applied magnetic field. This implies that $\RPol$ is also temperature independent, because, as far as spin polarization is concerned, an increase (a decrease) of temperature, at a fixed field, is equivalent to a decrease (an increase) of the magnetic field, at a fixed temperature.

For EuS, we took $B_0=0.007$~T, which is the magnetic field required to observe the SP experimentally \cite{apl2020}, and the same $B_0$ was taken for EuO. For EuSe, we took $B_0=0.07$~T, because it gives the best agreement between the experimentally measured peak SP magnetic moment, and the one obtained from Monte Carlo simulations in this work. The exact source of $B_0$ for EuSe is not clear, but it could be due to lattice spin fluctuations that cause modulation of the exchange interaction between lattice spins, in this strongly metamagnetic system \cite{kasuyaEuSe}.

\section{Monte Carlo simulations in $\mbox{EuX}$}
\label{sec:MC}
Monte Carlo simulations of SPs were done for EuX. The Eu spins ($S=7/2$), of effective magnetic moment $\mueff=\gS\muB\sqrt{S(S+1)}$ \cite{blundellbook}, were distributed on a face centered cubic lattice.
Born-Karm\'an periodic conditions were imposed on a cube, of edge $L$. The photoexcited hole was placed at an Eu atom at the center of the cube. To assure that the SP was fully contained in the cube, $L$ was at least twice the radius of the SP, i.e. $L\ge2\RPol$. The very large cube required in some cases make the MC simulations very lengthy. In this work, EuO required the largest $L$ (see Table~\ref{tab:pars}), and the MC simulations extended over several days.

If a magnetic field is present, no matter how small, then the result of the MC simulations should be independent of the initial distribution of lattice spins. In the MC simulaton for EuS, EuTe and EuSe shown here, the spins were initialized in different ways: antiferromagnetic, ferromagnetic, in a random orientation, and all converged to the same result. For a random initialization, the azimuth angle of the spin vector, $\varphi$, was associated with a random  value between 0 and $2\pi$, and the cosine of the polar angle, $\theta$, with a  random value between -1 and +1. A randomly oriented spin was obtained from two random numbers, $q_1$ and $q_2$, between 0 and 1, giving $\varphi=2\pi q_1$ and $\theta=\arccos(1-2q_2)$. To generate random numbers, we used the linear congruential random number generator (routine ran0), from Ref. \onlinecite{numrec}.

However, a ferromagnetic initialization is closer to an SP than a random one, hence ferromagnetic initialization speeds up the convergence of the MC simulations. For this reason, in our MC simulations for EuO, the lattice spins were initialized in the ferromagnetic order. The MC cube in EuO is much larger than in the other EuX, while the effective exchange field, $\BXf(r)$ of equation \eqref{eq:bxf}, is much smaller, due to a much larger SP volume (see Table~\ref{tab:pars}). These two factors lead to an unacceptably large computing time to achieve convergence for random spin initialization, while ferromagnetic initialization brings the convergence time down to a feasible level.

The energy of the $i$-th spin in the lattice, $E_i$, for a fixed orientation of all other lattice spins, was calculated by adding its exchange interaction energy with its first and second neighbors (described by constants $J_1$ and $J_2$, respectively), to its Zeeman energy \cite{prb2014}. The input parameters used in the MC simulations are given in Table~\ref{tab:pars}.

The spin reorientation at each lattice site was done iteratively, using the Metropolis algorithm \cite{MetropolisAlgorithm1953,numrec}. A new random orientation was considered for a lattice site $i$, and the corresponding energy $E'_i$ was calculated. According to Boltzmann statistics, the probability that the $i$-th spin switches to the new orientation, at a given temperature, is given by
\begin{equation} \label{eq:p_spinflip}
p=\frac{e^{-(E_i'-E_i)/\kB T}}{1+e^{-(E_i'-E_i)/\kB T}},
\end{equation}
where $\kB$ is the Boltzmann constant.
To reject or accept the new orientation, a random number $q$ between 0 and 1 was generated. When $q> p$, the new orientation was accepted, otherwise it was rejected.

This spin reorientation procedure was done for all lattice sites contained in the MC cube, and repeated 3000 times over the whole cube.

 At the end of the MC simulation, the magnetic moment of a sphere of radius $\RPol$ was computed in two situations: one in which the exchange interaction between the photoexcited electron and the lattice spins was switched on (i.e., $\BXf$ was switched on), and another in which it was switched off ($\BXf$ was switched off). The difference between the two magnetic moments yields the net magnetic moment of an SP.

The final temperature dependence of SP magnetic moment was obtained from the average of 50 calculated MC curves.

\begin{table*}[!ht]
\begin{ruledtabular}
\begin{tabular}{cccccccc}
EuX & $J_1$ (K) & $J_2$ (K)            & $\JXf S$ (eV)                & $a_B/a$              & $B_0$ (T) & $\RPol/a$ & $L/a$\\
    &       &                   &                            & from Ref. \ocite{apl2020} & &from eq. \eqref{eq:RPol} & \\
\hline
EuTe & 0.043 \cite{WachterP} & -0.15 \cite{WachterP} & 0.29 \cite{prb2014} & 1.84   & 0.1   & 4.2   & 10\\
EuSe & 0.290 \cite{prl2018} & -0.300 \cite{prl2018} & 0.263 [\onlinecite{prl2018}]& 2.68   & 0.07  & 4.9   & 10\\
EuS  & 0.221 \cite{eusJ1J2}      & -0.100 \cite{eusJ1J2} & 0.25 [\onlinecite{apl2020}]  & 3.28   & 0.007 & 8.98  & 18\\
EuO  & 0.606 \cite{mauger}      &  0.119 \cite{mauger} & 0.25  (this work)& 8.91   & 0.007 & 11.0  & 22\\
\end{tabular}
\end{ruledtabular}
     \caption{Input parameters used in the Monte Carlo simulations. The parameter source is indicated. The values of $B_0$ and $L/a$ are discussed in the main text.}
     \label{tab:pars}
\end{table*}

\begin{figure}
\includegraphics[angle=0,width=90mm]{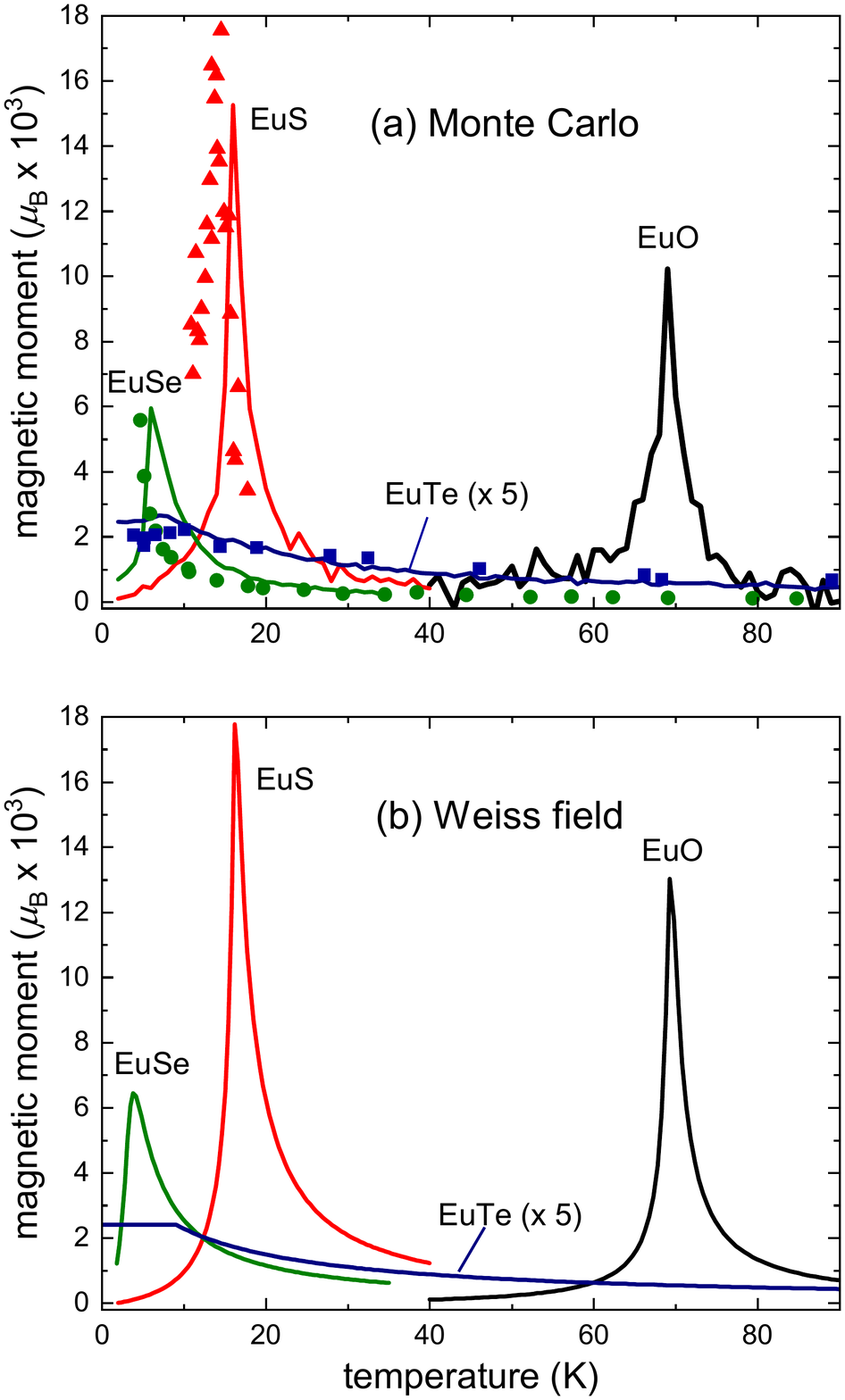}
\caption{(a) Results of the Monte Carlo simulations. Experimental results for EuTe (squares), EuSe (triangles) and EuS (circles) are also shown, from references [\onlinecite{prb17EuTe}], [\onlinecite{prl2018}], and [\onlinecite{apl2020}], respectively. (b) Results of the Weiss field model.
}
\label{fig:MC_WF}
\end{figure}

Figure~\ref{fig:MC_WF}~(a) depicts the SP magnetic moment for all EuX, obtained by the above described Monte Carlo simulations.
Available experimental data, taken from efs. \cite{prb16Rapid}, \cite{prl2018} and \cite{apl2020} for EuTe, EuSe and EuS, respectively, is also depiicted in figure ~\ref{fig:MC_WF}~(a). For EuO, experimental data is not yet available. Excellent agreement (within experimental error) of theory and experiment is achieved in the absolute maximum value of the SP size. The small difference between the theoretical and experimental temperature of the maximum is attributed to the specifics of the investigated samples, which were thin epitaxial layers grown by molecular, in which case residual strain may cause the critical temperature to differ slightly from the bulk \cite{Kepa,Beatriz2008} Inaccuracies in the bulk values of $J_1$ and $J_2$ (table \ref{tab:pars}) may also be at fault.

Figure~\ref{fig:MC_WF}~(a) shows that, for the ferromagnetic members EuO and EuS, $\muPol$ shows a very sharp peak at the Curie temperature, and the maximum $\muPol$ value is two orders of magnitude greater than in antiferromagnetic EuTe, whose $\muPol$ does not show a sharp maximum, but a plateau instead. The $\muPol$ temperature dependence for EuSe shows the same sharp maximum at the critical temperature, seen for  ferromagnetic EuO and EuS. EuSe is often categorized as  antiferromagnetic \cite{wachter,kasuyaEuSe}, however, applying a very small magnetic field at low temperatures changes the order to ferromagnetic \cite{kasuyaEuSe}. The effective exchange magnetic field produced by the photoexcited electron is sufficient to induce a ferromagnetic phase in the interior of the SP \cite{prl2018}, which is the region probed by the MC simulations. This is why EuSe displays the behavior characteristic of the ferromagnetic members of the EuX series.

The MC simulations displayed in Figure \ref{fig:MC_WF}~(a) show that above the critical temperature, $\muPol$ decreases for all EuX, because the SP is destroyed by thermal agitation of the lattice spins. When the interior of the SP is ferromagnetic,  ferromagnetic alignment onsets at $T<\TC$, therefore the photoexcited electron can increase the magnetization no further, and $\muPol$ vanishes. In contrast, when the interior of the SP is antiferromagnetic, $\muPol$ shows a plateau below $\TN$, at a value much less than the peak seen for a ferromagnet. These differences have at their root the antiferromagnetic exchange interaction between lattice spins, which prevents full spin polarization within the SP sphere.


The number of lattice spins involved in forming the SP increases with the cube of the ratio between the Bohr radius and the lattice parameter. From the Bohr radii given in Table~\ref{tab:pars}, in EuO the number of Eu spins contained in the SP is two orders of magnitude larger than in EuS. This makes it tempting to conclude that SPs in EuO could be much larger than in EuS \cite{apl2020}. However, the MC simulations indicate that the SP size in EuO and EuS are quite similar. This is because the Curie temperature of EuO is almost 5 times greater than for EuS (see Table~\ref{tab:parsWF}). Consequently, at the Curie temperature thermal quenching of lattice spin polarization works against the SP size much more efficiently in EuO than in EuS. Moreover, for a larger Bohr radius, due to the normalization of the wave function, the effective exchange magnetic field acting on an individual lattice spin becomes smaller, which reduces the radius of the SP sphere. As result, the maximum size SP that can be photogenerated in EuO and EuS end up being of similar size.

\section{Weiss field model}
In this section, the Weiss field (WF) applied to the description of photoinduced SPs is described separately for ferro and antiferromagnetic semiconductors (subsections \ref{subsec:FM} and \ref{subsec:AFM}, respectively). An application of the WF model to EuX and a comparison of the WF and MC models is left for subsection \ref{subsec:WFappl}.

In the Weiss field model, the spin vectors in every magnetic sublattice are substituted by an average. This approach allows to describe the Heisenberg exchange interaction by an effective magnetic field, the so-called molecular field, or Weiss field \cite{reifBook}. Because the interaction between individual spins is substituted by an average, spin fluctuations are not accounted for in the WF model. Nevertheless, the WF model describes remarkably well the main features of magnetism \cite{reifBook}(p.~433-434).

\subsection{Ferromagnetic case}
\label{subsec:FM}
In the ferromagnetic case, the average spin points in the direction of the magnetization, therefore the Weiss field associated with the Heisenberg exchange interaction becomes proportional to the magnetization. Then, the magnetization of the ferromagnet at a temperature $T$, and under an applied field $\bf B$, can be obtained from the transcendental equation
\begin{equation}
M(T)=\MSAT \Brillouin\left(\mu^*\frac{\bf B+\lambda \bf M}{\kB T}\right),
\label{eq:transc}
\end{equation}
where $\Brillouin$ represents the Brillouin function of order $S$, $\MSAT=N\mu^*$ is the saturation magnetization \cite{blundellbook}. The term $\lambda\bf M$ in equation~\eqref{eq:transc} represents the Weiss field, where $\lambda$ describes the exchange iinteraction between lattice spins.
By approximating the Brillouin function $\Brillouin(x)=\alpha_S x$, where $\alpha_S=\frac{S+1}{3S}$ in \eqref{eq:transc}, $\lambda$ is found to be connected to the Curie temperature of the ferromagnet, $\lambda=\frac{\kB\TC}{\alpha_S \MSAT\mu^*}$ \cite{blundell}.

The vector associated with $\BXf(r)$ is parallel to the applied magnetic field \cite{ijmp2009A}, hence the argument of eq.\eqref{eq:transc} becomes a scalar, giving:
\begin{equation}
M(r,T)=\MSAT\Brillouin\left(\mu^\ast\frac{\BXf(r)+B+\lambda M}{\kB T}\right).
\label{eq:transcXf}
\end{equation}

The magnetic moment of the SP was calculated as follows. Equation \eqref{eq:transcXf} was solved numerically for $M(r,T)$, using the Newton-Raphson method in one dimension (Ref.~\onlinecite{numrec}, sec. 9.4). $M(r,T)$ was integrated in the SP sphere, of radius $\RPol$ defined by equation \eqref{eq:bxf}, to give the total magnetic moment of the SP sphere. As always, the net magnetic moment of the SP was found by subtracting the magnetic moment of the same sphere, but calculated with $\BXf=0$.

The WF model can be used to produce a formula for the magnetization dependence on magnetic field at the Curie temperature in ferromagnets (see, for instance, Ref.~\onlinecite{blundellbook}, p.~90).  Here the same approach is used to find an analytical expression for the SP magnetic moment of the SP polaron at $T=\TC$. Expanding the Brillouin function in \eqref{eq:transc} in a power series, $\Brillouin(x)=\alpha_S x - \beta_S x^3$, where $\alpha_S=\frac{S+1}{3S}$ and $\beta_S=\frac{(S+1)^2+S^2}{90S^3}(S+1)$, and substituting $\lambda=\frac{\kB\TC}{\alpha_SN{\mu^*}^2}$, equation \eqref{eq:transc} can be resolved for the magnetization at $T=\TC$, at a field $B$:
\begin{equation} \label{eq:MTCofB}
M(\TC)=\MSAT\left(\frac{\alpha_S^4\mu^*\,B}{\beta_S \kB\TC}\right) ^{1/3}.
\end{equation}
Substituting $B$ in \eqref{eq:MTCofB} by $\BXf(r)$, as given by equation \eqref{eq:bxf}, and integrating \eqref{eq:MTCofB} in the SP sphere of radius $\RPol$, we obtain $\muPol$ at $T=\TC$:
\begin{equation}
\muPol=
27\,A\,\mu^\ast \left(N\aB^3\right)^{2/3}\left(\frac{ \pi^2 \alpha_S^4}{\beta_S}\right)^{1/3}
\left(\frac{\JXfS}{\kB \TC}\right)^{1/3}\,
\label{eq:muPol_Weiss}
\end{equation}
where
\begin{equation}
A=1-\left(\frac{x_0^2}{2}+x_0+1\right)e^{-x_0},
\label{eq:A}
\end{equation}
and $x_0=\frac{2\RPol}{3\aB}$.

\subsection{Antiferromagnetic case}
\label{subsec:AFM}
Here the photoinduced SP magnetic moment is derived analytically for an antiferromagnetic semiconductor, at any temperature.
We will approximate the antiferromagnetic system by dividing it into two magnetic sublattices, $\bm M_+$ and $\bm M_-$ respectively, of equal magnitude, but generally pointing in different directions.
For simplicity, only the exchange interaction between spins in different sublattices was included. The exchange interaction between spins on the same sublattice can be easily added, but has no significant effect \cite{yosida} and it is therefore ignored here. In this case, the sublattice magnetizations, $\bf M_+$ and $\bf M_-$, will obey the equation
\begin{equation}
\left|\bf{M}_\pm\right|=\frac{1}{2}\MSAT\,\Brillouin\left(\mu^\ast\frac{\left|\bf{B}-\lambda M_\mp\right|}{\kB T}\right),
\label{eq:AFM}
\end{equation}
where the factor $\frac{1}{2}$ accounts for the saturation magnetization of each sublattice to be only half of the total.

Below the N\'eel temperature $\TN$, in zero magnetic field, each sublattice becomes spontaneously magnetized,
although the total magnetic moment of the sample remains zero, since $\bf M_-=-\bf M_+$. Then from eq.~\eqref{eq:AFM} with $B=0$, proceeding as in section \ref{subsec:FM}, $\lambda$ in equation \eqref{eq:AFM} can be written in terms of the N\'eel temperature:
\begin{equation}
\lambda=\frac{2\kB\TN}{\alpha_S \MSAT\mu^\ast}.
\label{eq:lambdaAFM}
\end{equation}

If a magnetic field greater than the spin-flop field is applied, the sublattice magnetization vectors, $\bf M_+$ and $\bf M_-$, tilt towards the magnetic field direction, by the same angle $\theta$, so that the sample magnetization becomes
\begin{equation}
M(T\le \TN)=2\left|{\bf M}_\pm\right| \cos\theta.
\label{eq:magAFM}
\end{equation}
A simple analysis shows that for fields below saturation (see, for instance, Ref.~\onlinecite{yosida})
\begin{equation}
\cos\theta=\frac{B}{2\lambda\left|{\bf M}_\pm\right|},
\label{eq:cos}
\end{equation}
therefore from eq.~\eqref{eq:magAFM}, for $T\le\TN$ the magnetization is linear on the applied magnetic field, irrespective of the temperature:
\begin{equation}
M(T\le \TN)=\frac{B}{\lambda}.
\label{eq:belowTN}
\end{equation}

For $T>\TN$, $|\bf M_\pm|\rightarrow 0$, and for any applied field $\bf B$, we can approximate ${\bf M}_+={\bf M}_-$, and $\bf M\pm$ parallel to $\bf B$. Then, using \eqref{eq:AFM}, the transcendental equation for $|{ M}_\pm|$ becomes
\begin{equation}
{M}_\pm=\frac{1}{2} \MSAT \Brillouin\left(\mu^\ast\frac{B-\lambda M_\pm}{\kB T}\right).
\end{equation}
Using the linear approximation for the Brillouin function, which is valid for small arguments, we obtain $|{\bm M}_\pm|=\frac{\TN}{T+\TN}\frac{B}{\lambda}$. Therefore the total magnetization for $T\ge\TN$ becomes
\begin{equation}
M(T\ge \TN)=\frac{2\TN}{T+\TN}\frac{B}{\lambda}.
\label{eq:aboveTN}
\end{equation}
From equation \eqref{eq:cos}, the saturation magnetic field, $\BSAT$, which is required to impose complete ferromagnetic alignment at $T=0$~K is given by
\begin{equation}
\BSAT=\lambda \MSAT.
\label{eq:BSAT}
\end{equation}
With this in mind, equations \eqref{eq:belowTN} and \eqref{eq:aboveTN} can be united into a single expression, valid in fields below the saturation value:
\begin{equation}
M(T)=\MSAT\frac{B}{\BSAT}\left\{
\begin{array}{cc}
1 &\mbox{if $T\le \TN$}\\
\frac{2\TN}{T+\TN} & \mbox{if $T\ge \TN$} \end{array}
\right.
\label{MAFMvsT}
\end{equation}

Substituting $B$ by $\BXf(r)$ in equation \eqref{MAFMvsT}, and integrating in the SP sphere, i.e. $r\le\RPol$, and using equations \eqref{eq:lambdaAFM} and \eqref{eq:BSAT}, we find an analytical expression for the magnetic moment of a photoinduced SP in an antiferromagnet, as a function of temperature:
\begin{equation}
\frac{\muPol}{\muB}=\frac{3\JXfS}{2\kB\TN}
D\left\{
\begin{array}{cc}
1 &\mbox{if $T\le \TN$}\\
\frac{2\TN}{T+\TN} & \mbox{if $T\ge \TN$} \end{array}
\right. ,
\label{eq:muPolAFMvsT}
\end{equation}
where $D=1-\left(\frac{u_0^2}{2}+u_0+1\right)e^{-u_0}$, and $u_0=2\RPol/\aB$. This result is independent of $B$, as long as the saturation field is never reached within the SP sphere.

\subsection{The Weiss field model applied to SPs in EuX}
\label{subsec:WFappl}

In contrast to the Monte Carlo approach, using
the Weiss field model to describe SPs, requires deciding a priori whether the material within the SP sphere is ferro, or antiferromagnetic. The europium chalcogenides, EuO and EuS are natural ferromagnets, while EuSe and EuTe are antiferromagnets. However, for EuSe, a small magnetic field at low temperatures can induce a transition into the ferromagnetic phase \cite{kasuyaEuSe}. In our WF model for the SP, we treat EuSe using the ferromagnetic picture, because the exchange field is sufficient to make the interior of the SP ferromagnetic, as discussed in section \ref{sec:MC}.

To calculate the SP magnetic moment in the WF approximation, we used as input parameters $\JXfS$, $\aB$ and $\RPol$, given in Table~\ref{tab:pars}.
In addition to these three parameters, the WF model requires the critical temperature $\TC$ or $\TN$, totaling four input parameters. The values of critical temperatures are listed in Table~\ref{tab:parsWF}. We used $\TC=4.8$~K for EuSe, which is the temperature at which the measured SP magnetic moment displays a maximum \cite{prl2018}.

\begin{table}[!ht]
\begin{ruledtabular}
\begin{tabular}{ccc}
EuX & $\TC$ (K) & $\TN$ (K) \\
\hline
EuO &  69 \cite{mauger}     &   --      \\
EuS &  16 \cite{WachterP} & \\
EuSe & 4.8 \cite{prl2018} & \\
EuTe & -- & 9.6 \cite{eusJ1J2} \\
\end{tabular}
\end{ruledtabular}
\caption{Critical temperatures used in the Weiss field model.}
\label{tab:parsWF}
\end{table}

Figure ~\ref{fig:MC_WF}~(b) shows the WF model results for all EuX members. As can be observed in a comparison of the top and bottom panels of figure~\ref{fig:MC_WF}, the WF model reproduces the same qualitative behavior obtained by MC simulations. Quantitatively, the SP maximum values obtained by the WF model are about 10\% larger than those found from MC simulations. This small difference can be attributed to the complete absence of fluctuations in the WF model, which naturally contributes to enhancing the SP size. Minor differences in the critical temperatures can be attributed to uncertainties in the input parameters.

\section{Conclusions}
In conclusion, we presented here the first theory ever reported on the photoinduced SP magnetic moment dependence on temperature in magnetic semiconductors. The Monte Carlo simulations indicate that the SPs are most efficiently induced in ferromagnetic semiconductors, but only around the critical temperature, whereas in antiferromagnetic ones, SPs are efficiently induced at any temperature below the critical one.

Because the MC simulations are very demanding computationally, we developed a much simpler alternative model based on the  Weiss field theory, which reproduces the MC simulations very well. The simplicity of the WF model makes it accessible to a wide audience. The success of the highly simplified WF model, which is based on a simplified effective mass Hamiltonian, at all temperatures and in a whole family of materials, is quite remarkable, given that an SP is a complex many-body system, which involves many thousands of interacting particles.

The WF model produces analytical formulas that can be extended to other magnetic semiconductors. The much investigated GdN can be used as an example.  Stoichiometric GdN is reportedly antiferromagnetic \cite{wachterGdN2012}, hence
from formula \eqref{eq:muPolAFMvsT}, the maximum SP that can be induced in GdN is equal to $3\muB\JXfS/2\kB\TN$.
The N\'eel temperature of stoichiometric GdN is in the range 20-30~K \cite{wachterGdN2012}.
The band-lattice exchange interaction $\JXfS$ has been estimated to be 0.35~eV by \onlinecite{wachterGdN2012}, but a value as large as 1.24~eV has also been reported \cite{Sharma_GdN_2006}. Substituting these parameters in \eqref{eq:muPolAFMvsT}, the expected maximum photoinduced SP size in stoichiometric GdN is found to be in the range $\muPol/\muB\sim 170-1200$. This large range of values is due to the uncertainty in the material parameters. Thus, even for the much studied magnetic semiconductor GdN, there is still insufficient knowledge for an accurate prediction of photoinduced SP size, using the WF model developed in this work.



\section{Acknowledgments}
This work was funded by CNPq (Projects 303757/2018-3 and 420531/2018-1) and FAPESP (Projects 2019/02407-7 and 2019/12678-8).

\providecommand{\noopsort}[1]{}\providecommand{\singleletter}[1]{#1}%

\end{document}